\documentstyle{article}

\newcommand{\ba}{\begin{array}}
\newcommand{\ea}{\end{array}}
\newcommand{\beq}{\begin{equation}}
\newcommand{\eeq}{\end{equation}}

\begin{document}

\setlength{\topmargin}{-0.8cm}
\addtolength {\oddsidemargin} {-2.0cm}
\setlength{\parskip}{1pc}
\setlength{\parindent}{0pc}

\begin{center}
{\Large \bf Response of an Excitatory-Inhibitory Neural
Network to External Stimulation: An Application to Image Segmentation}

\vspace{0.5cm}
{\bf Sitabhra Sinha and Jayanta Basak}

Machine Intelligence Unit, Indian Statistical Institute,
Calcutta 700 035, India.\\
\end{center}

\begin{abstract}
Neural network models comprising elements which have exclusively excitatory
or inhibitory synapses are capable of a wide range of dynamic behavior,
including chaos. In this paper, a simple excitatory-inhibitory neural pair,
which forms the building block of larger networks, is subjected to
external stimulation. The response shows transition between various types
of dynamics, depending upon the magnitude of the stimulus. Coupling
such pairs over a local neighborhood in a two-dimensional plane,
the resultant network can achieve a
satisfactory segmentation of an image into ``object'' and
``background''. Results for synthetic and and ``real-life'' images 
are given. 
\end{abstract}

\section{Introduction}
Dynamical transitions in brain activity, in the presence of an external
stimulus, has received considerable attention recently. Most investigations
of these phenomena have focussed on the phase synchronization of
oscillatory activity in neural assemblies. An example is the detection
of synchronization of ``40 Hz'' oscillations within and between visual
areas and between cerebral hemispheres of cats \cite{Gra89} and other
animals. Assemblies of neurons have been observed to form and separate
depending on the stimulus.
This has led to the speculation that, phase synchronization of
oscillatory neural activity is the mechanism for ``visual
binding'' - the process by which local stimulus features of an object
(e.g. color, motion, shape), after being processed in parallel by different
(spatially separate) regions of the cortex are correctly integrated
in higher brain areas, forming a coherent representation (``gestalt'').

Sensory segmentation, the ability to pick out certain objects by
segregating them from their surroundings, is a prime
example of ``binding''. The problem of segmentation of
sensory input is of primary importance in several fields.
In the case of visual
perception, ``object-background'' discrimination is the most obvious
form of such sensory segmentation: the object to be attended to,
is segregated from the surrounding objects in the visual field. 
Several methods for segmentation, both classical \cite{Gon92,Pal93}
and connectionist (e.g., see \cite{Gho92}) are reported in literature.

Most of the studies on segmentation through neural assembly formation
has concentrated on networks of oscillators which synchronize
when representing the same object.
Malsburg and coworkers \cite{Mal92} have sought to explain
segmentation through {\em dynamic link architecture} with synapses
that rapidly switch their functional state. Similar approaches
using synaptic couplings which change rapidly depending on the
stimulus have been used in a neural model for segmentation by
Sporns {\em et al} \cite{Spo91}.  
Grossberg and Sommers \cite{Gro91} have performed figure-ground separation
with a network of oscillators, some of which belong to the ``object'' and
the others to the ``background''. Oscillations of the former are
synchronized, whereas the others have non-oscillatory activity.
Han et al \cite{Han98} have used an oscillatory network for
Hopfield-type auto-association in pattern segmentation, using
the temporal dynamics of the nonlinear oscillators driven by
noise and subthreshold periodic forcing. If the input is a
superposition of several overlapping stored patterns, the network
segments out each pattern successively, as synchronous activation of a
group of `neurons'. Similar segmentation through synchronization of
activity among a cluster of neurons have been shown by other groups
\cite{Som90,Sch94b,Wan95,Cam96}. In contrast to this approach, we
present a method of utilizing the transition between different types
of dynamics (e.g., between fixed-point and periodic behaviors) of the
network elements, for performing segmentation tasks.

In this paper, we investigate the dynamical response of an excitatory-
inhibitory network model (with autonomous chaotic behavior) to
external stimulation of constant intensity $I$ (in time). We focus on how
the behavior of an individual element within the network changes with
$I$. A theoretical analysis has been presented for the transition from
period-2 cycles to fixed-point behavior for an isolated excitatory-
inhibitory pair (i.e., not coupled to any other element). Simulation
results for the spatially interacting coupled network are presented.
The application of the system for segmenting gray-level images is
studied. Finally, the possible improvements of the proposed method
are discussed.
\section{The Excitatory-Inhibitory Network Model}
The model, we have based our investigations, comprise
excitatory and inhibitory neurons, coupled to each other over a
local neighborhood.
As we are updating the model only at discrete time intervals, it is
the time-averaged activity of the neurons that is being considered.
Therefore, the ``neuronal activity'' can be represented as a continuous
function, saturating to some maximum value, inversely related to the
absolute refractory period. This is arbitrarily taken to be 1, so that
we can choose the sigmoid function to be the neural activation function
for our model:
\beq
\ba{lll}
F_{\mu} (z) & = & 1 - \exp(-{\mu} z), ~~{\rm if}~~ z \geq 0\\
	    & = & 0, ~~{\rm otherwise}.
\ea
\label{refsgm}
\eeq
The basic module of the proposed network is a pair of excitatory and
inhibitory neurons coupled to each other (Fig. 1(a)).
If $x$ and $y$ be the activities of the excitatory and the
inhibitory elements respectively, then they evolve in time according to:
\beq
\ba{lll}
x_{n+1} & = & F_a(w_{xx} x_n - w_{xy} y_n + I_n)\\
y_{n+1} & = & F_b(w_{yx} x_n - w_{yy} y_n + I^{\prime}_n),
\ea
\label{refexin}
\eeq
where, $w_{ij}$ is the weight of synaptic coupling from element $j$
to element $i$, $F$ is the activation function defined by (\ref{refsgm}
and $I, I^{\prime}$ are external stimuli. By imposing the following
restriction on the values of the synaptic weights:
$$ \frac{w_{xy}}{w_{xx}} = \frac{w_{yy}}{w_{yx}} = k,$$
and absorbing $w_{xx}$ and $w_{yy}$ within $a$ and $b$ (respectively),
we can simplify the dynamics to that of the following one-dimensional
difference equation or ``map'':
\beq
z_{n+1} = F_a(z_n + I_n) - k F_b(z_n + I^{\prime}_n).
\label{refnmap}
\eeq
Without loss of generality, we can take $k = 1$. In the following
account we will be considering only time-invariant external stimuli,
so that, for our purposes:
$$ I_n = I^{\prime}_n = I.$$
The resultant neural map, exhibits a wide range of dynamics (fixed
point, periodic and chaotic), despite the
simplicity of the model \cite{Sin98c,Sin98f}.

\subsection{Dynamics of a single excitatory-inhibitory neural pair}
The autonomous behavior (i.e., $I,I^{\prime} = 0$)
of an isolated pair of excitatory-inhibitory
neurons show a transition from fixed point to periodic behavior
and chaos with the variation of the parameters $a, b$, following
the `period-doubling' route, universal to all unimodal maps \cite{Str94}.
The map
\beq
z_{n+1} = {\bf F}(z_n) = F_a(z_n) - F_b(z_n),
\label{refnmap1}
\eeq
describing the dynamics of the pair (Fig. 1(b)),
has two fixed points, at $z_1^* = 0$ and $z_2^*$
(which is the solution of
the transcendental equation $ z = \exp (-b z) - \exp (-az) $). The fixed
point, $z_1^*$ is stable if the local slope $ \simeq (a-b)$) is less than 1.
For $$a > \frac{1}{1-\mu},$$
where $\mu = \frac{b}{a}$, this condition no longer holds and $z_1^*$
loses stability while $z_2^*$ becomes stable, by a transcritical bifurcation.
On further increase of $a$ (say), this fixed point also loses stability,
with the local slope becoming less than -1, and a 2-period cycle occurs.
Increasing $a$ leads to a sequence of period-doublings ultimately giving
rise to chaos.   

The introduction of an external stimulus of magnitude $I$ has the
effect of horizontally displacing the map, Eqn. (\ref{refnmap1}),
to the left by $I$. This implies that $z_1^*=0$ is no longer a fixed
point, while the other fixed point $z_2^*$ is now a solution of the
equation $ z = \exp (-b (z+I)) - \exp (-a(z+I)) $. The slope at $z_2^*$
decreases with increasing $I$ - giving rise to a reverse period-doubling
transition from chaos to periodic cycles to finally, fixed-point behavior.
\subsection{Analysis of response to constant magnitude external stimulus}
We shall now consider how the dynamics of the excitatory-inhibitory pair
changes in response to external stimulus. Let us consider the isolated
neural pair, whose time evolution is given by Eqn. (\ref{refnmap}). On
replacing the expression of the transfer function from (\ref{refsgm}), we get
\beq
z_{n+1} = \exp(-b(z_n + I)) - \exp(-a(z_n + I)).
\label{ref1ee}
\eeq
Now,
$$z_{n+1} = z_n = z^{*}$$ for a fixed point. 
It is stable if 
$$ \frac{dz_{n + 1}}{dz_n} \geq -1,$$
i.e.,
$$(a - b)\exp(-a(z_n + I)) - bz_{n+1} \geq -1.$$
Therefore, for the fixed point to be marginally stable
(i.e. ${\frac{dz_{n+1}}{dz_n}} = -1$), it must satisfy the following
condition:
\beq
(a-b)\exp(-a(z^* + I_c)) = bz^* - 1,
\label{ref1q}
\eeq
where, $I_c$ is the critical external stimulus for which $z^*$ just
attains stability.

Let us define a new variable, $\alpha$, as
\beq
\alpha = \frac{b z^* - 1}{a - b}.
\label{ref2}
\eeq
Therefore, from (\ref{ref1q}), we get
\beq
\exp(-a( z^* + I_c )) = \alpha.
\label{ref3}
\eeq
Also from (\ref{ref2}),
\beq
z^* = \frac{1}{b} + \frac{1-\mu}{\mu}\alpha
\label{ref4}
\eeq
where $\mu = b/a$.
Now, from (\ref{ref1ee}), a marginally stable
fixed point can be expressed as 
$$z^* = -\exp(-a(z^* + I_c)) + \exp(-b(z^* + I_c)).$$
Therefore, from (\ref{ref3}) and (\ref{ref4}), the above expression
can be written as
\beq
\alpha^{\mu} - \alpha = \frac{1}{b} + (\frac{1 - \mu}{\mu})\alpha.
\eeq
By simple algebraic manipulation, we get
\beq
\alpha = \frac{1}{b^{1/\mu}}(1 + a\alpha)^{1/\mu}.
\label{addref}
\eeq
Assuming $a\alpha << 1$,
we need to consider only the first order terms in $\alpha$
in the right hand side, so that
\beq
b^{1/\mu}\alpha = 1 + \frac{a\alpha}{\mu},
\label{ref5}
\eeq
which gives the following expression for $\alpha$:
\beq
\alpha = \frac{1}{b^{1/\mu} - \frac{a}{\mu}}.
\label{ref6}
\eeq
For a real solution of $z^*$ to exist, we must have $bz^* - 1 > 0$, since,
otherwise, $z^*$ will have an imaginary component (from (\ref{ref1})). In
other words, $\alpha > 0$ (from (\ref{ref1q})).
Therefore, from (\ref{ref6}), we must have
\beq
a < \mu b^{1/\mu}.
\eeq
Since $b = \mu a$, we get 
\beq
a > \mu^{\frac{\mu + 1}{\mu - 1}}.
\label{refamu}
\eeq
For example, if $\mu = 0.5$ then $a > 8$ for $z^*$ to be real.
From (\ref{ref1q}) we get 
\beq
\exp(a I_c) = \frac{a-b}{bz^* -1}\exp(-az^*).
\label{ref1}
\eeq

Taking logarithms on both sides, we have,
$$I_c = -z^* - \frac{1}{a}\log(\alpha).$$
Therefore replacing $z^*$ from (\ref{ref4}),
\beq
I_c = (\frac{\mu - 1}{\mu})\alpha - \frac{1}{\mu a} - \frac{1}{a}\log(\alpha).
\label{ref7}
\eeq
The equation (\ref{ref7}), together with (\ref{ref6}), provides the
critical value of the external stimulus which leads the oscillatory neuron
pair to a fixed stable state, subject to the restriction (\ref{refamu}).

This expression can be further simplified.
From (\ref{addref}), we can write
$$\mu \log(\alpha) = -\log(b) + \log(1 + a\alpha).$$
As before, assuming $a\alpha << 1$, we need to consider only the first
order terms in $\alpha$ in the
right hand side of the logarithmic expansion, which gives us
\beq
\log(\alpha) = \frac{a\alpha}{\mu} - \frac{1}{\mu}\log(b).
\label{ref8}
\eeq
From (\ref{ref6}), (\ref{ref7}), and (\ref{ref8}), the critical magnitude of
the external stimulus is given as
\beq
I_c = \frac{1-\frac{2}{\mu}}{(\mu a)^{1/\mu} - \frac{a}{\mu}} + \frac{1}{\mu
a}\log(\frac{\mu a}{e}),
\label{maineqn}
\eeq
where $e = \exp(1)$. Fig. 2 shows the $a$ vs. $I_c$ curves for different
values of $\mu$, viz. $\mu$ = $0.1$, $0.25$ and $0.5$.

\subsection{Choosing the operating region}
To make the network segment regions of different intensities ($I_1 <
I_2$, say), one can fix $\mu$ and choose a suitable $a$, such that
$I_1 < I_c < I_2$. So elements, which receive input of intensity $I_1$,
will undergo oscillatory behavior, while elements receiving input of
intensity $I_2$, will go to a fixed-point solution. 
Notice that, the curves obtained from (\ref{maineqn}) gives two values of
$a$ for the same $I_c$. This gives rise to an
operational question: given a certain $I_c$, which value of $a$
is more appropriate? Notice that, the region of the $a$ vs $I_c$ curve
(Fig. (2))
to the left of the maxima, has a very high gradient. This implies that,
in the presence of wide variation in the possible value of $I_c$, choice
of $a$ from this region will show very small variation - i.e., the
system performance will be robust with respect to uncertainty in the
determination of the appropriate value of $I_c$.
This is possible in the case of any gray-level image with a bimodal
intensity distribution, having a long, almost uniform valley in between
the two maxima.

On the other hand, the region of the curve to the right of the maxima
has a very low gradient (almost approaching zero for high values of $a$).
This implies structural stability in network performance, as wide
variation in choice of $a$ will give almost identical results.
So, choice of $a$ from this region is going to make the network
performance stable against parametric variations.
As both robustness against uncertain input and stability against
parametric variations are highly desirable properties in network
computation, a trade-off seems to be involved here. The nature
of the task in hand is going to be the determining factor of
which value of $a$ we should choose for a specific $I_c$.
\section{The two-dimensional network of excitatory-inhibitory neural pairs}
The introduction of spatial interactions over a local
neighborhood in the above model produces some qualitative changes in the
response of the network to external stimulus.
We have considered discrete approximations
of circular neighborhoods \cite{Bis85} of
radii $r_{ex}, r_{in}$ ($r=1,2)$ in our simulations (Fig. 3(a)).
A theoretical study of the
changes in the dynamics due to spatial interactions is in progress and
will be reported later.

There is an important feature to consider about the relative
sizes of the neighborhoods
of the excitatory and inhibitory
neurons, ${\cal{R}}_{ex}$ and ${\cal{R}}_{in}$,
respectively. The autonomous dynamics of a coupled network is given by
$$z_{n+1} = {\bf F} (z_n + \frac{1}{|{\cal{R}}_{ex}|} \Sigma_{i \in 
{\cal{R}}_{ex}} x^i_n - \frac{1}{|{\cal{R}}_{in}|} \Sigma_{
i \in {\cal{R}}_{in}} y^i_n),$$
where $|{\cal{R}}|$ indicates the number of neurons within a neighborhood.
This can be rewritten as
$$z_{n+1} = {\bf F} (z_n + \frac{1}{|{\cal{R}}_{ex}|} \Sigma_{i \in 
{\cal{R}}_{ex}} z^i_n + \frac{1}{|{\cal{R}}_{ex}|} \Sigma_{i \in 
{\cal{R}}_{ex}} y^i_n - \frac{1}{|{\cal{R}}_{in}|} \Sigma_{i \in
{\cal{R}}_{in}} y^i_n).$$
If we take $r_{ex} = r_{in}$, then the equation reduces to
$$z_{n+1} = {\bf F} (z_n + \frac{1}{|{\cal{R}}_{ex}|} \Sigma_{
i \in {\cal{R}}_{ex}} z^i_n ).$$
As $z_n \geq 0$ at all sites $i$, the activation increases with $n$,
thereby driving the network to a homogeneous, uniformly activated state.

Therefore, unless $r_{ex} < r_{in}$, the
network activity becomes unstable owing to the unbounded increase
in the activity of the excitatory elements. This is seen by looking
at the averaged activity of the network, ${\overline{z}}_n = \frac{1}{N}
\Sigma_{i=1}^N z_n (i)$, where $z (i)$ indicates the $i$-th
excitatory-inhibitory neural pair. For $r_{ex} = r_{in}$,
${\overline{z}}$ shows an
oscillatory behavior whose amplitude increases with $n$ (Fig. 3(b)).
For stable behavior, the
amplitude of oscillation should be constant in time (this is so for $r_{ex} <
r_{in}$, as shown in Fig. 3(c)).
\section{Simulation and Results}
The response behavior of the excitatory-inhibitory neural pair, with
local couplings, has been utilized in segmenting images and the results
are shown in Figs. 4 and 5. Both synthetic and ``real-life'' gray-level
images have been used. The initial state of the network is taken to be
totally random. The image to be segmented is presented as external
input to the network, which undergoes 200 - 300 iterations.
Keeping $a$ fixed, a suitable value of $\mu$ is chosen from a
consideration of the histogram of the intensity distribution of the image.
This allows the choice of a value for the critical
intensity ($I_c$), such that,
the neurons corresponding to the `object' converge to fixed-point
behavior, while those belonging to the `background' undergo period-2
cycles. In practice, after the termination of the specified
number of iterations,
the neurons which remain unchanged
over successive iterations (within a tolerance value, $th$) are labeled
as the ``object'', the remaining being labeled the ``background''.

The synthetic image chosen is that of a square of
intensity $I_2$ (the ``object'')
against a background of intensity $I_1$ ($I_1 < I_2$).
Uniform noise of intensity $\epsilon$ is added to this image. The
signal-to-noise ratio (SNR) is defined as the ratio of the range of 
gray levels in the original image to the range of noise added (given
by $\epsilon$). 
For SNR = 1, the results of segmentation are shown in
Fig. 4. Fig. 4(a) shows the original image while segmentation
performance of the uncoupled network is presented in Fig. 4 (b). As is
clear from the figure, the isolated neurons perform poorly in
identifying the `background' in the presence of noise. 
The segmentation performance improves remarkably when spatial interactions
are included in the model.
Results for $r_{ex}=1, r_{in}=2$ and $r_{ex}=r_{in}=2$
are shown in Figs. 4(c) and (d), respectively.
The two architectures show very
similar segmentation results, at least upto the iterations considered here,
although the latter is unstable (as discussed in the previous section).
Excepting for the boundary of the `object',
which is somewhat broken,
the rest of the image has been assigned to the two
different classes quite accurately.

We have also considered the 5-bit gray level ``Lincoln'' image 
(Fig. 5(a)) as an
example of a ``real-life'' picture. A suitable $I_c$ has been estimated
by looking at the histogram of the gray-level values, and taking the
trough between two dominating peaks as the required value. As in
the synthetic image, the performance of a network of uncoupled neurons is
not satisfactory (Fig. 5(b)). The results of including spatial
interaction are shown in Figs. 5(c) and (d).
Most of the image has been labeled accurately, except
for a few regions (e.g., near the neck).

Note that, we have considered a single value of $a$ (and hence $I_c$)
for the entire image. This is akin to ``global thresholding''.
By implementing local thresholding and choice of $a$ on the basis of
local neighborhood information, the performance of the network
can be improved.
\section{Discussion}
In the present work, we have assumed constant connection weights
over a local neighborhood. However, a gaussian profile of weights
may be biologically more realistic. One can also make the critical
intensity $I_c$ proportional to the ambient intensity. This is
in tune with how the retina seems to alter its sensitivity to
incoming visual stimuli \cite{Wer72}. Finally, the role of variable
connection weights, that can be implemented in the present model
by changing the value of $k$ (ratio of the weights), may be
investigated.

Chaotic neurodynamics has been used to segment images by Hasegawa
{\em et al} \cite{Has96}. However, their method is based on
using chaos to avoid the local minima problem in the variable
shape block segmentation method.
We have instead concentrated on using stimulus induced transitions
in neural network dynamics to segment images. This is closer to
the neurobiological reality. As Malsburg \cite{Mal92} has indicated,
the reason oscillatory synchronization has been studied so far,
as a mean of segmenting sensory stimuli is its relative ease of
analysis. However, with the developments in nonlinear dynamics
and chaos theory, we can advance to segmentation using more
general dynamical behavior.

\section*{Acknowledgements}
We would like to thank Prof. S. K. Pal (MIU, ISI)
for his constant encouragement.

\vspace{1cm}
\begin{center}
\Large{Figure Captions}
\end{center}

{\bf Fig. 1} (a) The basic excitatory ($x$)-inhibitory ($y$) neural module,
and (b) the 1-dimensional neural map, ${\bf F}$ (Eqn. (\ref{refnmap1})),
with the activation functions for the constituent
excitatory (slope, $a = 20$) and inhibitory ($b = 5$) neurons.

{\bf Fig. 2} Critical magnitude ($I_c$) of the external stimulus, at which 
transition from periodic to fixed point behavior occurs. The circles 
(filled and blank) and squares represent the values obtained exactly 
through numerical procedures for $b/a = \mu =$ 0.5, 0.25 and 0.1, 
respectively. The curves indicate the theoretically predicted values.

{\bf Fig. 3} (a) Discrete circular neighborhoods of radii, $r = 1$
and $r = 2$.
Average activity (${\overline{z}}$) of a network of $100 \times 
100$ elements, arranged in a two-dimensional plane, with coupling over 
a local neighborhood: (b) $r_{ex} = 2$, $r_{in} = 2$ and (c) $r_{ex} = 1$, 
$r_{in} = 2$.

{\bf Fig. 4} Results of implementing the proposed segmentation method on 
noisy synthetic image: (a) original image, (b) output by the uncoupled 
network, (c) output by the coupled network ($r_{ex}=1, r_{in}=2$), 
and (d) output by the coupled network ($r_{ex}=r_{in}=2$), after 200 
iterations ($a$=20, $b/a$=0.25 and threshold $th$=0.02).

{\bf Fig. 5} Results of implementing the proposed segmentation method on 
``Lincoln'' image: (a) original image, (b) output by the uncoupled 
network, (c) output by the coupled network ($r_{ex}=1, r_{in}=2$), 
and (d) output by the coupled network ($r_{ex}=r_{in}=2$), after 300 
iterations ($a$=30, $b/a$=0.25 and threshold $th$=0.02).
\pagebreak

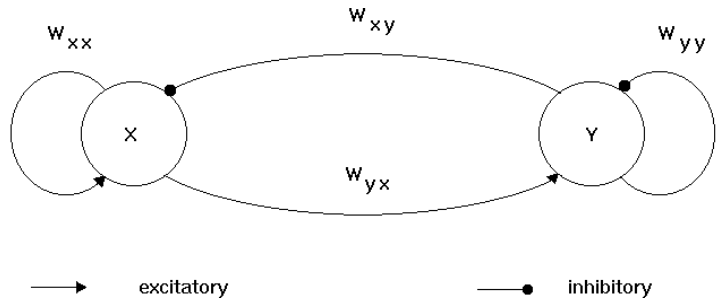
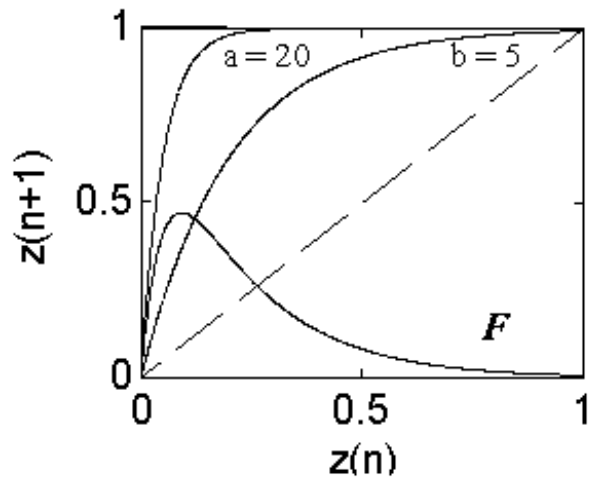
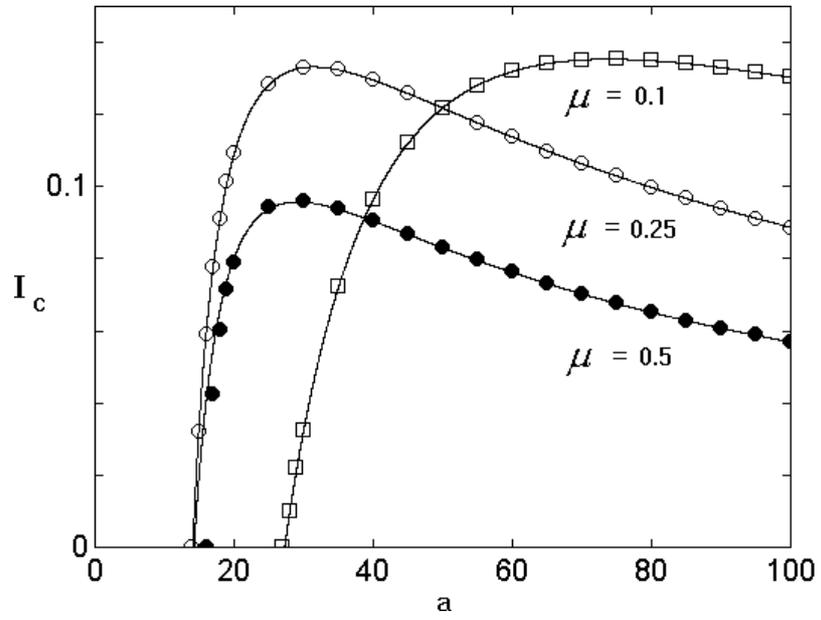
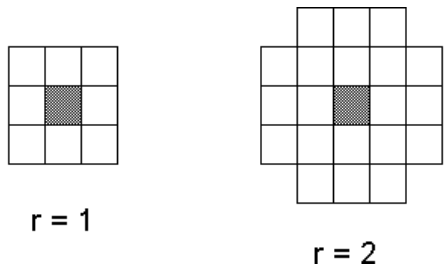
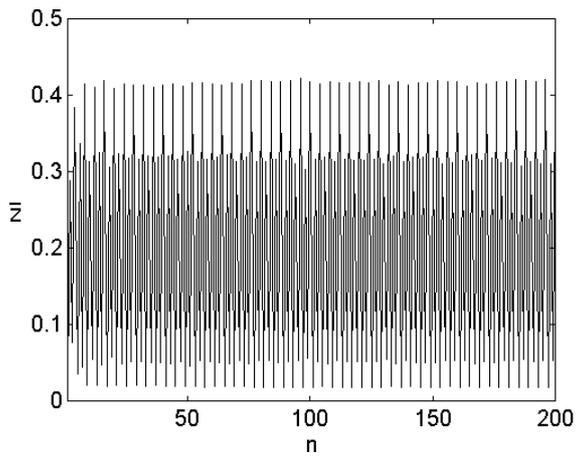
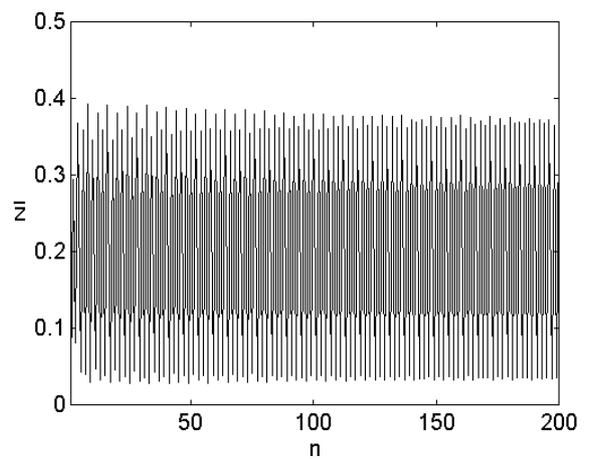
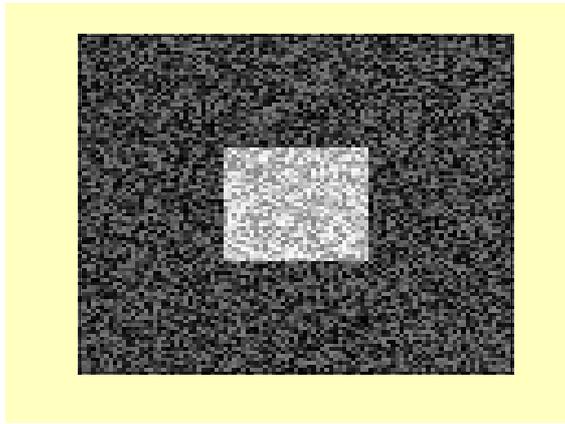
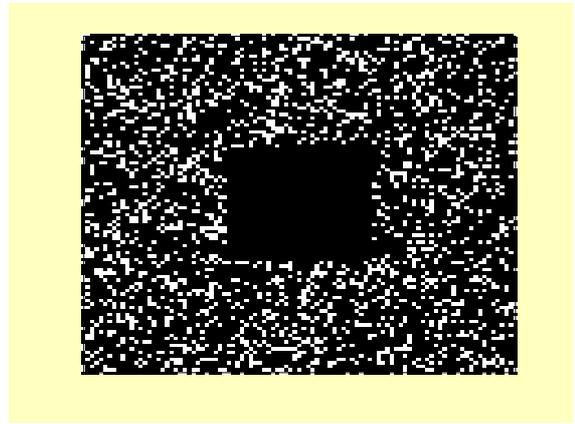
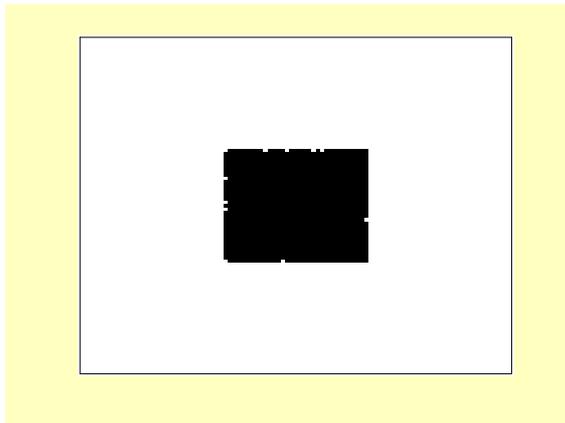
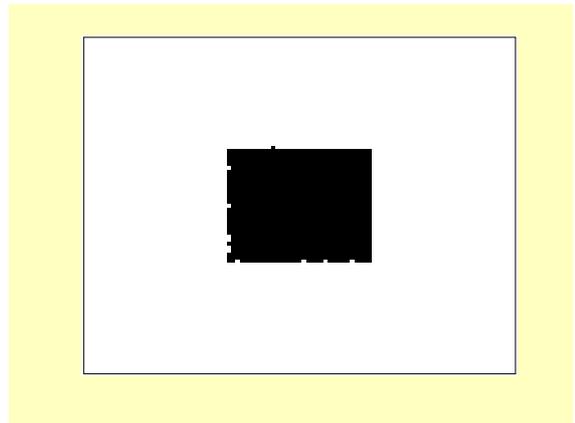
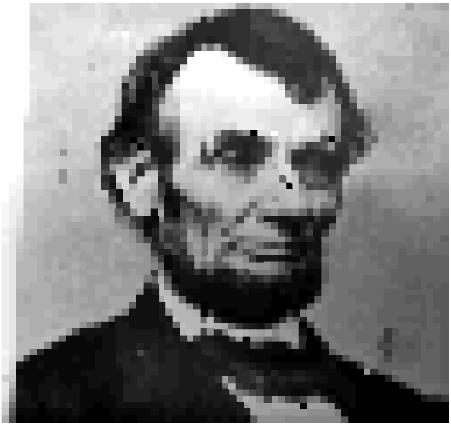
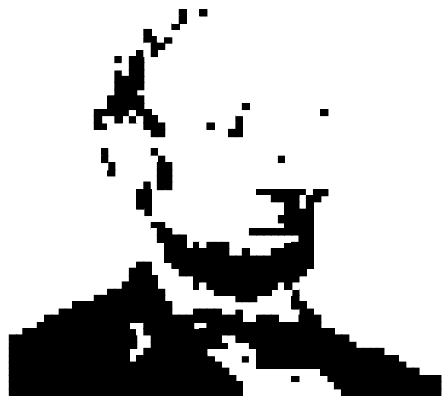
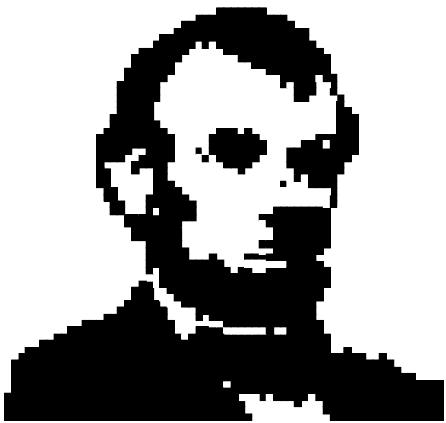
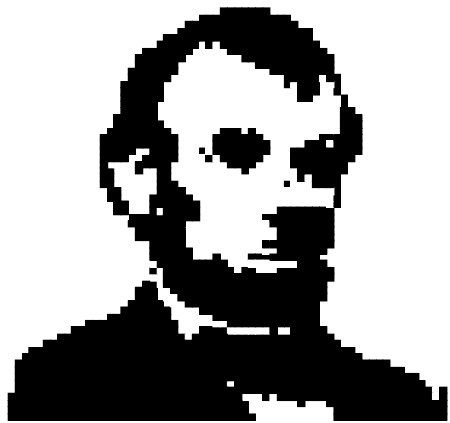
\end{document}